# Protein and ionic surfactants - promoters and inhibitors of contact line pinning

## Viatcheslav V. Berejnov[a]

**We report a new effect of surfactants in pinning a drop contact line, specifically that lysozyme promotes while lauryl sulfate inhibits pinning. We explain the pinning disparity assuming difference in wetting: the protein-laden drop wets a "clean" surface and the surfactant-laden drop wets an auto-precursored surface.**

To date, the effect of surfactants on a drop's wetting and spreading has been well established [1-12]. It was observed that low molecular weight surfactants extend spreading and decrease the contact line stability [13, 14]. However, the effect of surfactants and proteins on pinning a drop contact line has not yet been thoroughly appreciated despite being a primary issue for several important methods in life science. Optimization of the pinning conditions could benefit the crystallization of globular and membrane proteins [15, 16], formulation of the pesticide sprays for protecting the plants [1, 17], and the influence of the pulmonary surfactants on physiological wettability of alveoli in the lungs [18].

In this letter, we first study the quasi-static pinning of the protein and surfactant drops pinned by siliconized glass slides. We demonstrate that lysozyme (Lys) increases the hysteresis effect and stabilizes the drop contact line, enhancing the size of the completely pinned drops by a factor of 4-5 (compared to water). Conversely, the sodium dodecyl sulfate (SDS) reduces the apparent contact angles, decreasing the size of the completely pinned drops by a similar factor. We found the protein pinning to be similar to pinning induced by the geometrical and chemical corrugations of the contact line [15].

In our experiments, DI water was purified by NANOpure II (Barnstead, Boston, MA). Lys protein, 6-x times crystallized hen egg white, was purchased from Seikagaku America (Mr~14 kD, lot: LF1121, Falmouth, MA). Lys was dissolved in a 50 mM sodium acetate buffer with pH = 5. pH of the SDS (0.29 kD, 99%, Sigma) solutions was 9. All solutions were filtered.

Flat, siliconized 22 mm glass slides HR3-231 were purchased from Hampton Research (Laguna Niguel, CA). On a new siliconized slide, a water drop with a volume of ~20 μL formed a reproducible contact angle of ~(92±1)°. The siliconized material of the HR3-231 slides is similar to the organosilane-composed solution AquaSil (Hampton Research). We inspected the surface topography for some of the HR3-231 slides using a contact mode AFM (DI MultiMode III, Santa Barbara, CA) with NSC 1215 tips from MikroMasch. We found the manufacturer's coating to be homogeneous. We treated some hydrophobic slides chemically [15] in order to obtain the highly hydrophilic (contact angle <2°) circular area. Both the edge and hydrophilic-hydrophobic gradient

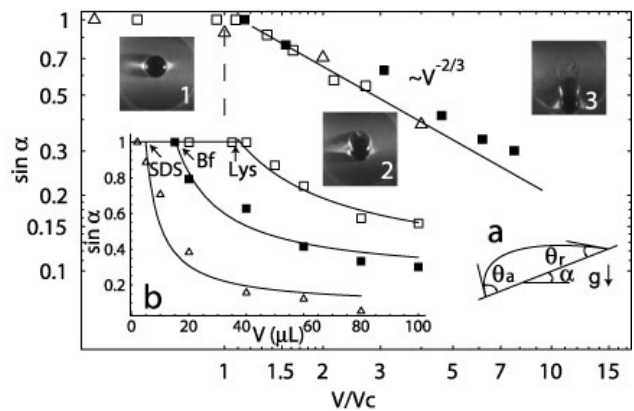

**Fig.1** Dimensionless diagram of the stability of inclined drops. The symbols: ■, □, and Δ correspond to Lys (5 mg/mL), buffer (Bf), and SDS (33.3 mg/mL) solutions, respectively. Insert (a) is a geometrical sketch of an inclined drop. Insert (b) subtracts $V_c$ from the raw-data diagrams (sin(α), V). The images 1, 2, and 3 represent three pinning zones: 1 - an absolutely stable drop; 2 - a drop is stable up to certain inclination < 90°, depending on V; and 3 - a drop is unstable and moves continuously.

provide a high threshold of pinning [15] that was taken as a reference in our experiments.

Each drop was dispensed manually onto initially horizontal glass slides, which were then slowly rotated in 2-4° steps (Fig. 1a). For the Lys drops the relaxation time between rotations was ~1 min to allow the transient disturbances of the drop to dissipate. For the SDS drops the initial spreading was similar to that described earlier [7-10, 12] and the relaxation time interval was different depending on concentration. We did not observe the autophobic [14] and Marangoni-induced [13] contact line displacements. The apparent advancing $\theta_a$ and receding $\theta_r$ contact angles were measured for some drops at quasi-static equilibrium using a horizontal goniometer [19]. Dispensed drop volumes were accurate to 0.1-0.5%, and tilt and contact angle measurements were accurate to 1-2°. The surface tension γ for the Lys solutions was measured using the pendant drop counting method [15].

We characterize pinning by measuring the critical tilt α corresponding to continuous motion of an inclined drop of volume V [19]. We observed that the scaling law $\sin(\alpha) \sim V^{-2/3}$ for large α may fit the (V, α) data for the different surfactant concentration C over a broad range of the supercriticality ratio $V/V_c$ (Fig.1). $V_c$ is a result of fitting, depends on C, and denotes the critical volume corresponding to a vertically (sin(α)=1) pinned drop (Fig.1b). The function (C, $V_c$) characterizes the surfactant pinning (Fig.2).

[a] *IESVic, University of Victoria, Victoria, BC V8W 3P6, Canada*
*E-mail: berejnov@gmail.com*

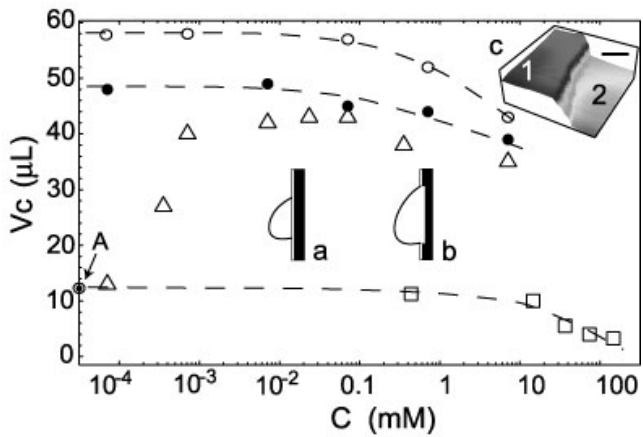

**Fig.2** Diagram of critical volumes versus concentration for vertically pinned drops. The inserts (a) and (b) represent drops on the untreated △ - Lys and □ - SDS, and treated ○ - Lys 8.0-diam-mm and ● - Lys 7.0-diam-mm glass slides, respectively. The point A depicts $V_c$ for water and buffer. (c) is an AFM image of the etched-2 versus unetched-1 areas for the treated slide (bar is 1μ). Dashed lines are not fits.

Two different regimes of pinning can be seen in Fig.2 depending on the chemical nature of the used surfactants. For very low concentrations close to point A, pinning of the Lys and SDS-laden drops behaves similarly. In the moderate and high concentration regions, adding Lys (triangles) and SDS (squares) increases and decreases the critical volumes of the pinned drops, respectively. The white and black circles represent the Lys drops initially dispensed on the circularly treated hydrophilic patterns with different radii [15] having edge profiles presented in Fig.2c. These patterns do not affect pinning of the SDS-laden drop. Both the similarities between the natural and pattern-induced pinning for the Lys drops and the differences in pinning between Lys and SDS are very remarkable.

Summarized below are our results corresponding to the different pinning regimes presented in Fig.2. First, pinning is associated with the appearance of a trace of liquid film behind the drop [20-22] that can be treated as a sign of high adhesion. Verifying this, we mark intervals on the concentration axis in Fig.3a and Fig.3b where the trace behind the drop body appears. The buffer and pure water do not exhibit any traces on the hydrophobic slide (Fig.4a), but the treated slides show unstable traces covering the treated area (Fig.4c). The Lys and SDS-laden drops show similar traces (Fig.4b and Fig.4d). However, high pinning was observed only for the Lys drops (Fig3a and Fig3b). The SDS decreases pinning in the trace-region. Thus, the presence of the trace does not correlate with the strength of pinning for our choice of surfactants.

Second, an equilibrium of an inclined drop yields the formula $\rho V_c g \sin(\alpha) \sim D\gamma\Delta(\cos\theta)_{r,a}$ [19, 20, 23], where $\rho$, $g$, and $\gamma$ are the density of liquid, gravity, and the surface tension, respectively, and the contact angle difference $\Delta(\cos\theta)_{r,a}$ equals $\cos\theta_r - \cos\theta_a$. Thus, the critical volume of an inclined drop (or pinning) is proportional to the wetted diameter D. Roughly, we can treat D as the diameter of an initially dispensed drop onto a horizontal substrate. This approximation holds in small drops when deformation of the drop surface describing by the Bond number $Bo=\rho g D^2/\gamma$ ($\lesssim 1$ for our case) is not too large. For the concentration range presented in Fig.3a we did not find a noticeable variation in D for the naturally dispensed Lys drops for either inclination or concentration. For the treated slides presented in Fig.2, increasing D qualitatively agrees with the above formula for low deviations of D in the Lys drops.

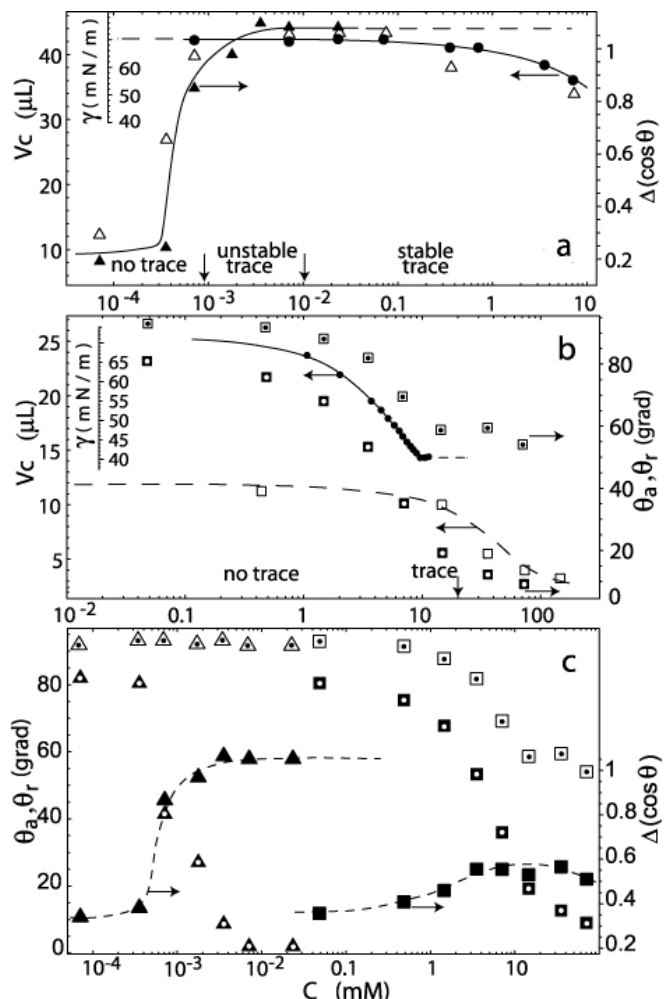

**Fig.3** Combined diagrams of critical volumes, contact angles, their difference, and surface tension versus the component concentrations. Black circles are $\gamma$, (a): Lys and (b): SDS [24]. △, ▲, □, and ■ denote $V_c$ and $\Delta(\cos\theta)_{r,a}$ for the Lys and SDS drops, respectively. Dotted triangles (Lys) and squares (SDS) depict the contact angles: white - maximal $\theta_a$ and black - minimal $\theta_r$. The zones of a liquid film trace appearing behind the quasi-statically displaced drop are marked on (a) and (b). Lines are not fits.

The patterns with larger radii (open circle) provide higher drop stability (larger pinning). The diameters of the SDS drops on hydrophobic slides grow monotonically with concentration [7]. However, unlike the Lys, the SDS decreases the drop pinning with respect to the concentration. Therefore, the D variation does not affect pinning for the SDS case, but it does for the Lys.

Third, we found that high pinning corresponds to high $\theta_a$ and $\Delta(\cos\theta)_{r,a}$, (Fig.3a and Fig.3c). For high concentration regions these parameters are related because the liquid-trace minimizes $\theta_r$ (Fig.3c dotted black triangles and squares). Keeping $\theta_a$ large is the only way to keep the difference $\Delta(\cos\theta)_{r,a}$ also large. Adding Lys to the drops does not affect the initially high values of $\theta_a$, and, consequently, $\Delta(\cos\theta)_{r,a}$ (Fig.3a and Fig.3c). Conversely, adding the SDS decreases $\theta_a$, controlling the overall low value of $\Delta(\cos\theta)_{r,a}$ adequately, and provides the low pinning effect.

Forth, the overall effect of concentration is noticeably different between Lys and SDS. Fig.3a demonstrates that pinning is tied to the Lys amount in the drop. Starting at $10^{-4}$ mM, the Lys-pinning increases as $\Delta(\cos\theta)_{r,a}$ increases, reaches its maximum corresponding to the minimum of $\theta_r$ (appearance of the liquid-

**2**

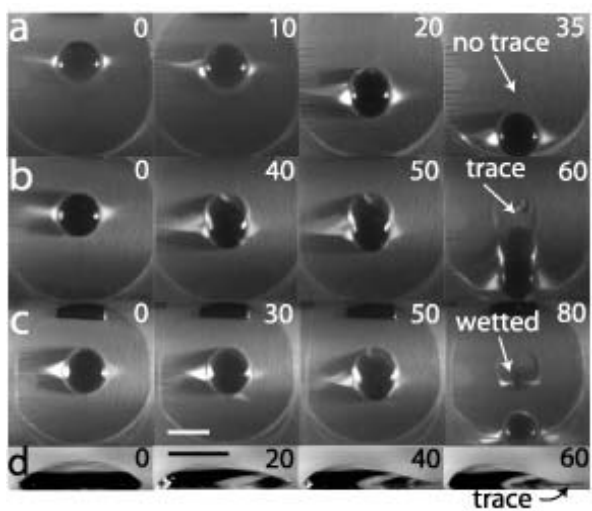

**Fig.4** Different regimes of displacement; (a): 40 μL drop of 50mM acetate buffer at pH=5 (water acts similarly); (b): 40 μL drop of 5mg/mL Lys; (c): 40 μl drop of water (bar is 5.5 mm); (d): side view of the 20 μL SDS 100 mg/mL drop (bar is 3.4 mm). Numbers are inclination in degrees; the slides (a), (b), and (d) are untreated, (c) is treated with D=5.5 mm.

trace), and finally saturates. Next, the high concentration region pinning decreases as $\gamma(C)$ decreases ($d\gamma/dC<0$). The SDS-pinning, however, is much less sensitive to concentration. For the SDS drops, $V_c$ is close to the water drop, is stable up to 10mM, and decreases as $\theta_r$ decreases. Additional probing of the Lys and SDS solutions with the plain glass slides (HR3-227, Hampton Research) does not show a qualitatively different behavior of pinning. For plain glass, the $V_c(C)$ curves generally follow the data presented in Fig.2, corresponding to the hydrophobic untreated slides for both the Lys and SDS surfactants.

We see that the appearance of a trace indicating the high adhesion between solution and glass surface does not provide high pinning unambiguously: Lys increases and SDS decreases pinning. Neither does extending the drop's wetted diameter correlate with a pinning increase: for Lys, buffer, and water, a variation in D due to patterning increases pinning while it does not for SDS. Pinning and contact angle difference are correlated as well: an increase in $\Delta(\cos\theta)_{r,a}$ corresponds to a pinning increase and decrease for the Lys and SDS solutions, respectively.

We attribute such a dramatic difference in pinning to the different gas-solid surfaces near the contact line. We hypothesize that the Lys contact line advances the "clean" [10] solid surface, whereas the SDS contact line advances the pre-cursored zone existing in close proximity to the contact line. We expect that mobility of the SDS molecules through a liquid interface may affect the advancing zone. Indeed, such mobility of the small amphiphilic molecules is well documented [2-4, 7-10, 12]. Though the mobility of proteins is not available, we argue it to be negligible compared to SDS. The globular proteins are weak surfactants with large, less amphiphilic, and therefore much less interfacially mobile molecules [25-27] and, in addition, they irreversibly adhere to the glass. We assume that the proteins adhere to the solid from the drop interior and provide high heterogeneity for the three-phase contact that results in the high pinning threshold similar to the corrugated case [15] (Fig.2, $C>10^{-2}$mM). In contrast, SDS penetrates the drop-solid-air interface, increases affinity between the solid and the drop, and thus decreases the pinning threshold.

The author thanks N. S. Husseini for proofreading the manuscript.